\documentclass[12pt,preprint]{aastex}

\newcommand{\citepeg}[1]{\citep[{e.g.,}][]{#1}}

\def\gr{\hbox{ \raisebox{-1.0mm}{$\stackrel{>}{\sim}$} }}

\shorttitle{GRB Afterglows}
\shortauthors{Zeh, Klose, Kann}

\begin{document}

\title{GRB afterglow light curves in the pre-\emph{Swift} era --
       a statistical study}

\author{A. Zeh,\altaffilmark{1}
S. Klose,\altaffilmark{1}
D. A. Kann\altaffilmark{1}}

\altaffiltext{1}{Th\"uringer Landessternwarte Tautenburg, Sternwarte 5,
D--07778 Tautenburg, Germany}

\begin{abstract}
We present the results of a systematic analysis of the world sample of
optical/near-infrared afterglow light curves observed in the pre-\emph{Swift}
era by the end of 2004. After selecting the best observed 16 afterglows with
well-sampled light curves that can be described by a Beuermann equation, we
explore the parameter space of the light curve parameters and physical
quantities related to them. In addition, we search for correlations
between these parameters and the corresponding gamma-ray data, and
we use our data set to look for a fine structure in the light curves.
\end{abstract}


\section{Introduction}

Nearly ten years after the discovery of the first Gamma-Ray Burst (GRB)
afterglow  in the optical/near-infrared \citep{Groot1997, vanParadijs1997},
the progress in the understanding of the GRB phenomenon is enormous
\citep[e.g.,][]{ZhangMeszaros2004, Piran2005}. From the observational point of
view nearly every individual afterglow has turned out to be specific in some
sense. The richness in afterglow properties observed so far is not much
surprising, however, given the fact that the afterglow phenomenon combines
internal properties of the underlying burster population with  properties of
the external circumburster medium. This in combination with redshift provides
a large parameter space for various flavors of GRB afterglows. However, in
spite of their individualities, as an ensemble afterglows trace the underlying
physical boundary conditions and the parameter space of the physical processes
involved. Revealing this parameter space is of fundamental interest since it
might much improve the understanding of the afterglow phenomenon. This has
already motivated several groups to perform a systematic analysis of
observational data from various afterglows \citep[e.g.,][]{Frontera2000,
PK2001a, PK2001b, PK2002, Yost2003, Frail2003a, Frail2004, Stratta2004,
Gendre2005, P2005a, P2005b}. The present work continues this kind of
investigations but concentrates on the optical/NIR bands.

While a fit of the observed broad-band spectral energy distribution (SED) of
an afterglow from the radio to the X-ray band is the best way in order to
extract the underlying physical parameters \citep[e.g.,][]{Frail2003b,
Yost2002, P2005b},  one might expect that the optical/NIR data alone are
completely explainable within the context of the underlying theoretical
model. In particular, in most cases it is the optical/NIR region where the
sampling and the quality  of the afterglow data is best.  It is clear that any
reliable theoretical model must be able to explain the observed richness in
the phenomenology of optical light curves of GRB afterglows. In this respect
it is worth summarizing and exploring the available data base on  afterglows
in the optical/NIR bands. This holds in particular with respect to the new era
in GRB research initiated by the launch and operation of the \emph{Swift}
satellite \citep{Gehrels2004}.

In the present paper we continue our systematic analysis of afterglow
parameters based on optical and near-infrared data (Zeh, Klose, \& Hartmann
2004, hereafter Paper I; Zeh, Klose, \& Hartmann 2005). Our study includes
$all$ afterglows in the pre-\emph{Swift} era with sufficient published
data. While in Paper I we analyzed the afterglow light curves with special
emphasis on an underlying supernova component, the goal of the present paper
is to explore the parameter space of the light curve parameters and of
physical quantities related to them. In Paper III \citep{KKZ}, we expand this
analysis to the spectral energy distribution of afterglows in the optical and
near-infrared bands in order to search for signatures of dust in GRB host
galaxies.

\section{Data analysis \label{data}}

We collected from the literature all available photometric data on GRB
afterglows observed by the end of 2004 and checked them for consistency. We
modeled the light curve of the optical transient (OT) following a GRB as a
composite of genuine afterglow light, supernova (SN) light, and constant light
from the underlying host galaxy. The afterglow is either modeled by a single
or a smoothly broken double power-law according to  \cite{Beuermann1999}.
Our fitting procedure is based on a $\chi^2$ minimization with a
Levenberg-Marquardt iteration. This minimization technique provides formal
uncertainties of the fitted parameters through the covariance matrix. For the
remainder of this paper, all uncertainties are 1$\sigma.$

We always fitted photometric magnitudes with a fitting equation as follows:
\begin{eqnarray}
m_{\rm OT}(t) &=& -2.5\,\log\{10^{-0.4\,m_c}
[(t/t_{\rm b})^{\alpha_1\,n} + (t/t_{\rm b})^{\alpha_2\,n}]^{-1/n} \nonumber \\
              &+& k\,10^{-0.4\,m_{\rm SN}(t/s)} + 10^{-0.4\,m_{\rm host}}\}\,.
\label{mag1}
\end{eqnarray}
The parameters in Equation~(\ref{mag1}) are the prebreak decay slope
$\alpha_1$, the postbreak decay slope $\alpha_2$, the break time $t_{\rm b}$,
the steepness of the break $n$, the brightness of the host galaxy $m_{\rm
host}$, the constant $m_c$ which corresponds to the magnitude of the fitted
light curve for the case $n=\infty$ at the break time $t_{\rm b}$ (it stands
for the intersection point of the prebreak and the postbreak slope,
without considering a smooth transition), as well as
the supernova parameters $k$ and $s$, which indicate the luminosity ratio and
the stretch factor normalized to SN 1998bw (cf. Paper I). If there is no break
in the light curve, then Equation~(\ref{mag1}) reduces to
\begin{eqnarray}
m_{\rm OT}(t) &=& -2.5\,\log\{10^{-0.4\,m_1}t^{\alpha} \nonumber \\
              &+& k\,10^{-0.4\,m_{\rm SN}(t/s)} + 10^{-0.4\,m_{\rm host}}\}\,,
\label{mag2}
\end{eqnarray}
where $m_1$ is the brightness of the afterglow at $t=1$ day after the
burst. Before the fitting process the data were corrected for Galactic
extinction using the maps of \cite{Schlegel1998}. For a more detailed
description of our procedure we refer to Paper I.

In the present paper, our input list contains 59 optical/NIR afterglows
observed in the pre-\emph{Swift} era which allowed us to perform a fit
(Table~\ref{tabres}).  Sixteen of these have a well-sampled light curve in at
least one photometric band (mostly the $R_C$ band). These light curves show a
well-detectable break and have sufficient data points before and after the
break time, so that the prebreak decay slope and the postbreak decay slope
according to Equation (\ref{mag1}) are well-defined. More precisely, our
selection criterium for the best-defined afterglow light curves is that the
$1\sigma$ error  is less than 0.2 for $\alpha_1$, less than 0.3 for
$\alpha_2$, and less than 1 day for $t_b$ (Table~\ref{tabres}).  We did not
use the accuracy of the fit ($\chi^2$ per degrees of freedom) as a selection
criterium, however.  We consider Equation (\ref{mag1}) as an empirical first
order approximation of an observed light curve and all deviations from the
corresponding fit as fine structure (\S~\ref{FineS}).  Light curves with no
break are by definition excluded from this sample of best-defined  afterglows,
because it cannot be ascertained with complete certainty if the slope is a
pre- or postbreak.  Whenever we did not  detect a break, the data quality was
usually insufficient to exclude the possibility of a jet break in the light
curve. In these cases there is either no early time data available, or the
break could have been missed because of a bright host galaxy or an underlying
SN component.

All but one (GRB 980519) of the 16 afterglows  in our sample have a known
redshift, none of these bursts is classified as an X-ray flash\footnote{While
GRB 030429 is classified as an XRF in the observer frame \citep{Sakamoto2005},
it would be an X-ray rich GRB in the host frame with $E_{peak}=93^{+32}_{-21}$
keV.}. For the present study the afterglows of GRB 021004 and GRB 030329 were
excluded from our analysis because of their many rebrightening episodes (see
also \S~\ref{FineS}). In the case of the afterglow of GRB 030329, we performed
two additional fits that are also given in Table \ref{tabres}, the details are
in Appendix \ref{individual}. While these fits give very good results, they
are based on only a small part of the light curve and we thus do not include
their results in the selected sample for statistical study, with the exception
of Figure \ref{a1n}. The afterglow of GRB 000301C also shows a rebrightening
episode \citep[around 3.5 days after the burst;][]{GLS2000, Gaudi2001}, but as
the deviations are not that large and occur only during a certain period, the
light curve can still be fitted with a broken power-law, even though
$\chi^2_{\rm d.o.f.}$ becomes relatively large.

A special note is required for GRB 021211. The afterglow light curve  of this
burst fulfills the aforementioned criteria for the amount of the $1\sigma$
error bars of the fit parameters, nevertheless we have not included it  in our
subsample of well-defined light curves. According to our data, the light curve
shows a break 0.11$\pm$0.09  days after the burst with a steepening by
$\Delta\alpha = \alpha_2 - \alpha_1 = 0.26 \pm 0.11$.  This is in close
agreement with the amount of steepening expected for the passage of the
cooling break across the optical window \citep[for a Compton parameter less
than 1; e.g.,][]{PK2001a}. In addition, \cite{Nysewander05} report on evidence
for color changes  of the afterglow just  around this time. Given the fact
that  no optical data have been reported in the literature for the time period
between  1 and 10 days after the burst, we consider it as very likely that the
break in the light curve we have found does indeed signal the passage of the
cooling break \citep[as already suspected by][based on their finding of color
changes]{Nysewander05}, while the real jet break occurred between 1 and 10
days after the burst. Note that the light curve break around 0.11 days is not
identical to the break discussed by others concerning the very early light
curve, which has been attributed to the reverse shock \citep{Li03,Fox03,
Wei03, Holland04, KP03}. Given these findings, and  since we try to keep our
subsample as homogeneous as possible, we have not included this burst in the
following study.

The light curve parameters we have deduced from our data might be slightly
different to those obtained and used by other groups.  This is mainly due to
the fact that we use a different, and most likely larger data base. In
addition, there is still some bias in the selection of the data, in the
definition of what are outliers, which data should be used and which data
should not. While we do not claim that our data base is the best for every
individual GRB, most probably it is the most comprehensive set. The strength
of our approach is that we analyze all afterglow light curves using the same
numerical procedure. In this respect we are confident that the results we
obtain are statistically robust.

\section{Results}

The results of our light curve fitting of the individual afterglows are
summarized in Table~\ref{tabres}. Here, the second column indicates the sample
of the best-defined 16 afterglows. In some cases we could only derive upper or
lower limits for some parameters, because the light curve was poorly sampled
(see \S~\ref{individual}). In the following we discuss only the results
obtained for this sample of 16 afterglows.

The distributions of the deduced light curve parameters are shown in
Figures~\ref{a1Histo} to \ref{logtbz}. The mean of the prebreak decay slope is
$\alpha_1 = 1.05\pm 0.10$, with $\alpha_1$ ranging from $0.58\pm0.05$ (GRB
000301C) to $\alpha_1=1.76\pm0.05$ (GRB 011121), while $\alpha_2$ ranges from
$1.30\pm0.02$ (GRB 041006) to $3.03\pm0.27$ (GRB 030429) with the mean at
$2.12\pm0.14$. About half of the postbreak decay slopes have $\alpha_2<2.0$.
Both distributions overlap in the interval from about 1.3 to 1.7. The
distribution of $\alpha_2$ is almost constant, with a possible cutoff around
1.3 and no preference for any value.

Figure~\ref{DaHisto} displays the distribution of the difference of the decay
slopes before and after the break, $\Delta\alpha=\alpha_2-\alpha_1$.  The
distribution is asymmetric with its maximum around 0.8 and a longer tail
towards higher values. It is notably broader than the distribution of
$\alpha_1$ and $\alpha_2$. It also indicates the possibility that some
afterglows could have very shallow breaks with $\Delta\alpha< 0.3$ that could
easily be missed. The gap around $\Delta\alpha=1.7$ is probably due to
low-number statistics and we do not consider it to be significant. In this
respect we cannot confirm the potential evidence for a bimodality of the
distribution of $\Delta\alpha$ \citep{P2005a}, allthough we cannot
reject this possibility either.

In Figure~\ref{logtbz} we present the distribution of the observed break times
for the afterglows with the best-defined light curves, but translated into the
corresponding GRB host frame. This distribution is strongly asymmetric with a
clear peak in the host frame at lower values around 0.3 days. That most breaks
occur at relatively early times supports the view that in several cases
(Table~\ref{tabres}) light curve breaks might have been missed due to a lack
of early-time data. The afterglow with the earliest break in the host frame
was GRB 041006, while the afterglow of GRB 000301C had the latest break
time. On the other hand, late breaks might have been missed in several cases
too, because the afterglow was already too faint at the break time, and/or an
underlying SN component or a bright host galaxy simply made the discovery of
the break in the optical bands impossible. We conclude that these data
indicate, even though they do not prove, that in fact all afterglow light
curves have detectable breaks due to a collimated explosion, as long as they
are not hidden by rebrightening episodes as in GRB 030329
\citep[e.g.,][]{Lipkin2004}. It is clear that any model that explains the
observed light curve breaks must be able to reproduce this observed
distribution (Figure~\ref{logtbz}).

The distribution of the shape parameter $n$ (Equation~\ref{mag1}) is more
difficult to quantify. While for about half of the afterglows the data allowed
us to determine $n$ during the fitting procedure, in the other cases $n$ did
not converge during the fit ($n\rightarrow\infty$), because the sampling of
the data is not good enough. It is worth noting that whenever we were able to
determine $n$, we obtained a relatively soft break ($n\approx1-2$) and in each
case the prebreak decay slope $\alpha_1$ was very shallow. In the
other cases we had to fix $n$ at a relatively large value in order to obtain
an acceptable fit, and we chose $n=10$. Choosing $n\approx1-2$ made the fit
worse. Whether this indicates a possible bimodal distribution of the parameter
$n$ is an open question. The distribution of the shape parameter $n$ is the
biggest unknown, so far, including its theoretical interpretation.

\section{Discussion}

\subsection{Wind versus ISM models \label{profile}}

Figure~\ref{a1a2} shows the correlation between the parameters $\alpha_1$ and
$\alpha_2$ in comparison with eight standard afterglow models that cover the
cases (1) $\nu < \nu_c$, (2) $\nu_c < \nu, Y<1$, (3) $\nu_c < \nu, Y>1,
2<p<3$,  (4) $\nu_c < \nu, Y>1, p>3$  for the ISM and for the wind model,
where $\nu$ is the observed frequency and  $Y$ is the Compton parameter
\citep[][their eqs. 21, 22 and 29]{PK2001a}.  Those models with $\nu < \nu_c$
and $Y<1$  require $p<2$ in order to explain an observed $\alpha_2<2$.  In
particular, there is a group of five bursts (GRBs 990123, 991216, 010222,
030328, 041006) that cluster around $\alpha_2=1.5$. Within the
corresponding 1$\sigma$ error bars $\alpha_2>2$ is basically ruled out and so
is $p>2$. Such shallow postbreak decay slopes cannot be explained by a flat
electron distribution either \citep {Dai2001, B2001, Wu2004}.

Based on the underlying theoretical models, which predict $p=\alpha_2$,
then we find that the parameter space of $p$ is rather broad, ranging from
about 1.5 to 3. While the results obtained for $p$ for the individual
afterglows differ among various authors, all studies agree that within the
current theoretical framework no evidence for a universality of $p$ is found
\citep[e.g.,][;Paper III]{PK2001a, PK2001b, PK2002, Preece2002, Yost2003,
P2005b}. This is contrary to what one might expect from theoretical models of
highly relativistic shocks \citep{Achterberg2001, Kirk2000}, and contrary to
what one might prefer on theoretical grounds \citep{FW2001}.

If one allows for $p<2$ then an inspection of Figure~\ref{a1a2} shows that a
wind model with $\nu < \nu_c$ is preferred for GRBs 980519, 990123, 991216,
000926, 020124, 020405, 030328, and 041006, even though GRBs 990123, 991216
and 000926 lie fairly off the models because of their relatively small
$\alpha_2$ or large $\alpha_1$. On the other hand, an ISM model with $\nu <
\nu_c$ is preferred for GRB 990510 and GRB 011211. GRB 020813 is either an ISM
case with $\nu < \nu_c$ or an ISM/wind case with $\nu > \nu_c$, in the
following, we will assume this to be an ISM case. Two afterglows (GRB 010222
and GRB 011121) could be an ISM or a wind case with $\nu > \nu_c$. In the case
of GRB 011121, the error bars are so large that $\nu < \nu_c$ cannot be fully
excluded.  Finally, the afterglows of GRBs 000301C, 030226, and 030429 are
outliers because of their relatively large $\alpha_2$. It is noteworthy that
all afterglows with soft breaks in the light
curve ($n\approx1-2$) belong to the group of bursts that are less compatible
with a wind profile, and all these afterglows have $\alpha_1<1$.

It is difficult to quantify whether the outliers really represent a different
population or whether in these cases the light curves are simply ill-defined.
For example,  the afterglow of GRB 000301C was affected by a strong
rebrightening episode that has been modeled by a gravitational microlensing
event \citep{GLS2000}. As these authors note, the removal of this event
leads to $\alpha_1\approx1.1$, which would shift GRB 000301C toward the
theoretical prediction of the ISM model with $\nu < \nu_c$. In the case of GRB
030226 there is evidence that the afterglow light curve showed fluctuations.
In combination with the relatively sparse set of postbreak data it is well
possible that the late-time observations stopped when the afterglow underwent
a fluctuation, so that finally  the deduced postbreak decay slope is too
large. On the other hand, early-time spectra of this afterglow reveal features
that can be best understood as due to a stellar wind profile
\citep{Klose2004}. In the case of 030429, \cite{Jakobssonz030429} find that
the light curve undergoes a significant rebrightening around 1.7 days after
the burst. They suggest to exclude the data around the rebrightening from the
fit and fix $\alpha_2=1.7$ (deduced from the SED and the $\alpha - \beta$
relations). In this case the light curve would be compatible with an ISM/wind
model and $\nu>\nu_c$, although it is unclear how large the error in
$\alpha_2$ would be.

If we neglect GRBs 000301C, 030226, and 030429, then Figure \ref{a1a2} shows
that the group of optical afterglows that is compatible with a wind model is
notably larger than the group of afterglows that prefers an ISM model.  While
basically all studies in the literature  agree that afterglows seem to
separate into a group that can be best described by a wind model and a group
that can be best described by an ISM model, our data show that the wind
scenario is statistically preferred. In fact, within their 1$\sigma$ error
bars nearly all $(\alpha_1, \alpha_2)$ pairs are compatible with a wind
profile while for the ISM model  such a statement is clearly ruled out.

\subsection{The jet opening angles \label{theta}}

Figure~\ref{thetaHisto} displays the distribution of the jet half-opening
angle, $\theta_{\rm jet}$, for our sample of 16 bursts (GRB 980519 is not
included as no redshift of this burst is known) as derived from the observed
break time, assuming the uniform jet model \citep{Rhoads1999}. We calculated
$\theta_{\rm jet}$ following \cite{Sari1999} for an ISM medium (GRBs 990510,
000301C, 011211, 020813, 030226, 030429) and \cite{Bloom2003} for a wind-like
medium (GRBs 990123, 991216, 000926, 010222, 011121, 020124, 020405, 030328,
041006), according to the results obtained for the density profile of the
individual afterglows (\S~\ref{profile}). For the isotropic equivalent energy
$E_{\rm iso}$, the radiation efficiency and the redshift we adopted the values
given by \cite{Friedman2005}. In the case of an ISM model we used the
circumburster density as given in \cite{Friedman2005}, while for the wind
model we assumed a mass-loss rate to wind speed ratio of $A_* =1$
\citep[cf.][]{Chevalier2000}. The distribution of $\theta_{\rm jet}$ that we
have found is strongly asymmetric with a peak between 2 and 5 degrees, has a
lower cutoff around 2 degrees and rapidly falls towards larger angles, in
agreement with what has been found in previous studies \citep{Frail2001,
PK2001b, PK2002, Bloom2003}.

\subsection{Correlations \label{correlations}}

Using the derived jet half-opening angles (\S~\ref{theta}) we find that the
distribution of the beaming corrected energy release in the gamma-ray band
ranges from $E_{\rm cor}=10^{49.9}$ erg (GRB 041006) to $E_{\rm
cor}=10^{51.4}$ erg (GRB 990123). In combination with the corresponding peak
energies, $E_{\rm peak}$, in the gamma-ray band \citep{Friedman2005}, in
Figure~\ref{EcorEpeak} we plot the correlation between $E_{\rm cor}$ and
$E_{\rm peak}$ in the GRB host frame, as it was  first reported by
\cite{Ghirlanda2004}. Considering  bursts \#9 and \#12 as outliers and
excluding them from the fit, we find $E_{\rm peak}\simeq 748\, (E_{\rm
51,cor})^\eta$, with $\eta = 0.79 \pm 0.09$.  This is  in qualitative
agreement with Ghirlanda et al. (2004) as well as \cite{Dai04}. On the other
hand,  there are differences, in particular concerning the existence of the
two outliers. They can be understood, however, since the fit includes
assumptions about the gas density in the GRB environment, which enters  the
calculation of the jet opening angle. We made use of the values provided by
\cite{Friedman2005} and these are higher by a factor of 10/3 compared to the
values used by Ghirlanda et al. (2004).  In addition, in several cases the
gamma-ray data given in \cite{Friedman2005} are notably different from those
used by Ghirlanda et al. (2004). Additionally, we also considered a wind-like
circumburst medium for some cases to calculate the jet half-opening angle,
while Ghirlanda et al. (2004) only regard an ISM-like circumburst medium.  It
is therefore not surprising that we do not exactly reproduce their results
(for a discussion, see also Friedman \& Bloom 2005). For instance, if we
reduce the assumed circumburster gas density  for bursts \#9 and \#12 from 10
to 1 cm$^{-3}$, then the corresponding data points do not fall out of the
sample anymore. A fit then provides $\eta = 0.78 \pm 0.09$.  While our data
base is too small in order to investigate the role of  outliers in the
Ghirlanda relation, we note that also the relation between the isotropic
equivalent energy release in the gamma-ray band and the intrinsic peak energy
\citep{Amati2002} is not very tight. In a recent study on BATSE GRBs,
\cite{Nak05} find that about 25\% of all bursts do substantially deviate from
this empirical relation \citep[see also][]{BP2005}.

In addition to the Ghirlanda relation, we have searched for linear
correlations between all individual afterglow parameters and between the burst
parameters in the gamma-ray band and the corresponding afterglow
parameters. Table~\ref{tabcor} lists the corresponding correlation
coefficients derived from weighted linear fits. The Ghirlanda relation
(here including the two outliers discussed above) is between $\log(E_{\rm
cor})$ and $\log(E_{\rm peak})$. Relatively tight correlations between
$\theta_{\rm jet}$ and $t_b/(1+z)$ and between $\theta_{\rm jet}$ and $\log
E_{\rm cor}$ are expected, as $\theta_{\rm jet}$ derives from $t_b/(1+z)$ and
$\log E_{\rm cor}$ derives from $\theta_{\rm jet}$. Note that the correlations
for the wind model are much tighter than for the ISM model, giving further
significance to the statistical conclusion that most circumburst environments
are wind-blown (\S~\ref{profile}). Next to the Ghirlanda relation and the
others just discussed, there are more correlations that seem significant
according to the absolute correlation coefficient, but visual inspection does
not support this conclusion. This holds also for the potential correlation
between  $\theta_{\rm jet}$ and $\alpha_1$ which has been reported for X-ray
data \citep{Liang2004}. Still, it is interesting to note that $\alpha_2$ seems
more or less correlated with all other parameters, including the redshift. On
the other hand, $\theta_{\rm jet}$, $\log E_{\rm cor}$ and $\log E_{\rm peak}$
are completely uncorrelated with the redshift. We conclude that no
evolutionary effect in the initial  explosion parameters is evident over a
wide range of redshifts.

Of special interest is the  parameter $n$,
which indicates the smoothness of the break (Equation~\ref{mag1}).  Even
though we could only determine $n$ for a few afterglows, it looks suspicious
that each time we had to fix $n$ to a relatively high value in order to get an
acceptable fit, the prebreak decay slope is $\alpha_1>1$. To test if this
could be due to a numerical problem we reconsidered the afterglow light curve
of GRB 030329 and fitted it only between 0.28 and 1 days, which includes the
time around the supposed jet break \citepeg{Uemura2003} but excludes the
cooling break \citep{Sato2003} and the rebrightening episodes
\citep[e.g.,][]{Lipkin2004}. For this time period the data density is high
enough in order to deduce a value for $n$ even if the break is very
sharp. Indeed, in this case we find  $\alpha_1=1.17\pm0.01, t_b=0.68\pm0.02$ days,
$\alpha_2=2.21\pm0.07$ and $n=7.54\pm1.47$. Adding this to our sample of afterglows
with a deduced parameter $n$ (Table~\ref{tabres}), a weak trend between $n$
and $\alpha_1$ becomes apparent (Fig.~\ref{a1n}). It indicates that a shallow
prebreak decline leads to  a smooth break or, seen the other way around, a
smooth break (a "rollover") implies a shallow prebreak decay slope. As we only
have very few values of $n$, we do not regard this as strong statistical
evidence for a correlation, since an  observational bias cannot be
excluded. Afterglows with a shallow decay are bright for a longer period of
time, which makes them easier to follow. Therefore, the data density around
the break  time is usually  higher compared to most afterglows with a steep
decline. A high data density around the break time is essential to determine
a value for $n$, however.  Nevertheless, it is worth to check if this trend is
confirmed in the \emph{Swift}-era.

Unfortunately, most bursts in our sample are at high redshift, so that no
supernova data are available. Consequently, no statistically founded
conclusions can be drawn on a potential correlation between afterglow
parameters and the corresponding supernova parameters (Paper I).

\subsection{Fine structure in the light curves \label{FineS}}

In principle, Equation (\ref{mag1}) represents a first order approximation of
an observed afterglow light curve. We consider any residuals that remain after
subtraction of the  fit from the observational data as the fine structure in
the light curve. Since the detection of fine structure in optical afterglows
depends  strongly on the sampling density of the light curve and the quality
of the data,  an observational bias cannot be excluded, which makes it
difficult to compare  the fine structure among the individual afterglows. On
the other hand, it is well-known that some afterglows show basically little or
no evidence for fine structure when sampled very densely with the same
telescope \citep[the best example being GRB 020813;][]{LaursenStanek2003},
while others show a  substantial amount of fine structure \citep[e.g., GRB
030329;][their figure 4]{Lipkin2004}. In order to investigate the occurrence
of fine structure in more detail, we have shifted all residuals to a common
evolutionary phase of an afterglow. We favor the idea that this can be done by
normalizing the time $t$ that has elapsed sind the GRB trigger to the break
time $t_b$ of the corresponding afterglow (Table~\ref{tabres}).  While
Figure~\ref{GS1} displays the individual light curves and corresponding
residuals that remain after subtraction of the fitted curves from the
observational data of the 16 afterglows in our sample, Figure~\ref{residuals}
displays all residuals in a single plot as a function of $t/t_b$.  We have
included here only those data with less or equal 0.05 mag individual
photometric error. The ratio $t/t_b$ is independent of redshift and allows us
to draw some general conclusions about the occurrence of fine structure in
afterglow light curves.

First at all, again  we find no evolutionary effect in the data. The  width of
the magnitude distribution of the fine structure of all 16 afterglows in the
prebreak evolutionary era spanning one decade in time ($0.1 \le t/t_b \le 1$),
is identical to the width of the  magnitude distribution in the postbreak era
spanning one decade in time ($1 \le t/t_b \le 10$), namely $\pm$ 0.1
magnitudes. The handful of data points around $t/t_b=0.2$ that reach beyond
$\pm$0.2 magnitudes mainly belong to GRB 011211 \citep[cf.][]{HollandAJ2002,
JakobssonNA2004} and is statistically not significant. This picture does not
change if we allow for larger individual photometric errors but then the
statistical significance of this finding becomes less strong. We conclude
that, on average, a patchy surface structure of afterglow shock fronts
\citep{Meszaros1998, Nakar2003} is not present in the photometry at times
later than $t/t_b>0.1$, with the probable exception of GRB 011211. In
addition, we find that the amplitude of the fine structure of all 16
afterglows as a group ($\pm$0.1 mag) is smaller by a factor of 4 compared to
the fine structure (or fluctuations) seen in the  optical afterglows of GRB
021004 \citep[e.g.,][]{deUga05} and GRB 030329 \citep{Lipkin2004}, which are
plotted in comparison\footnote{Note that our broken power law fits find late
breaks for both these afterglows (Table~\ref{tabres}), so most of the data is
at $t/t_b<1$.}. In other words, the latter two optical afterglows are indeed
very different from all other afterglows we have investigated, as the
deviations persist even after the break.

On the other hand, the residuals of the early afterglows of GRB 021004 and GRB
030329 are very similar. We find via $\chi^2$ minimization that a shift  of
the GRB 021004 light curve in $t/t_b$ by a factor of about  2.7 superposes the
early light curve evolution of both bursts (the $\chi^2$ minimum we find is
not sharp, shifts between 2.4 and 3.0 are acceptable). This is astounding, as
the bursts happened at two very different cosmological epochs. A deeper
analysis of this result will be pursued in a future publication (Kann et
al. 2005, in preparation).

\section{Summary}

Based on a systematic analysis of the optical and NIR data of \emph{all} GRB
afterglows with sufficient published data in the pre-\emph{Swift} era we have
explored the parameter space of the afterglow light curves and of physical
quantities related to them. From the 59 afterglows investigated
(Table~\ref{tabres}) we constructed a sample of 16 bursts with the
best-defined light curves useful for our purposes. Thereby we excluded the
afterglows of GRB 021004 and GRB 030329 because of their many re-brightening
episodes which made it difficult to fit them.

Using the sample of the 16 afterglows with the best defined light curves, we
find that in the optical bands, the average afterglow light curve is
characterized by a prebreak decay slope $\alpha_1 = 1.0\pm 0.1$ and a
postbreak decay slope  $\alpha_2 = 2.1\pm 0.1$. The distribution of both
parameters is rather broad but possible cutoffs are apparent in the available
data set. In particular, there is no evidence for a universality of $\alpha_2$
as it has been  predicted in some afterglow models. The distribution of the
break time in the host frame rises sharply towards smaller values, with the
most likely value at $t_b/(1+z)= 0.3\pm 0.2$ days.

We have then used the deduced light curve parameters to extract information
about the nature of the GRB environment using standard afterglow models
\citep{PK2001a}. We find that in most if not all cases the data are in
agreement with a wind model. A general preference of an ISM model is ruled
out. In addition, we find that the distribution of  the power-law index $p$ of
the electron distribution function  is rather broad ranging from about 1.5 to
3, supporting the view of a non-universality of $p$, in agreement with other
studies \citep[e.g.,][Paper III]{PK2001a, PK2001b, PK2002, Preece2002,
Yost2003, P2005b}.  Furthermore, we have searched in our data set for potential
correlations between the various light curve parameters and those that
characterize the corresponding burst in the gamma-ray band. With the exception
of the Ghirlanda relation between the beaming corrected energy release in the
gamma-ray band and peak energy in the GRB host frame (Ghirlanda et al. 2004),
no other tight correlation has been found. An intriguing correlation may exist
between the pre-break decay slope $\alpha_1$ and the smoothness parameter of
the break $n$, but more data is needed to verify this.

Finally we have analyzed in which manner the data indicate to a  general fine
structure that is superimposed a light curve decay according to the empirical
Beuermann double power-law. When normalized to the  corresponding break time
$t_b$ of a burst we find no evidence in the data that there is more structure
in the light curves  at times $0.1 < t/t_b < 1$ than at times $1 < t/t_b <
10$.  On the other hand, we find that the afterglows of GRB 021004 and
GRB 030329 are very different from all 16 afterglows in our sample. While the
latter vary  on average by only 0.1 magnitudes around the fitted light curve,
the former vary by 0.4 magnitudes. Moreover, the fine structure of
the light curves of GRB 021004 and GRB 030329 are initially very similar.

It is clear that more afterglows with well-sampled optical light curves are
needed in order to get better insight onto the parameter space of the physical
processes involved.  Even in the \emph{Swift} era this might not be an easy
task since most afterglows are simply very faint some days after the
burst. The observational challenge therefore remains the availability of
observing time on large optical telescopes in order to determine the light
curve parameters as well as possible.


\acknowledgments A.Z. and S.K. acknowledge financial support by DFG grant Kl
766/11-1. D.A.K. and S.K. acknowledge financial support by DFG grant Kl
766/13-2. We wish to thank S. Covino, J. Gorosabel, T. Kawabata,  B. C. Lee,
K. Lindsay, E. Maiorano, N. Masetti, R. Sato, P. M. Vreeswijk, and K. Wiersema
for contributing unpublished or otherwise unavailable data to the database,
and S. Cortes for reducing additional data. D.A.K. wishes to thank N. Masetti
and D. Malesani for enlightening discussions. We wish to thank S.  Barthelmy,
NASA, for the upkeep of the GCN Circulars and J. Greiner, Garching, for the
"GRB Big Table". We thank the anonymous referee for very constructive
remarks.


\appendix

\section{Notes on individual bursts \label{individual} }

\paragraph{GRB 970508}
The afterglow light curve of this GRB is anomalous, featuring an early plateau
phase enduring until one day after, followed by a steep ($\alpha\approx-3.4$)
re-brightening. Starting at 1.9 days, the afterglow decays with a simple power
law. Our fit starts at this point. The light curve is well sampled and no
break is seen, thus it is possible that the decay is postbreak (the break
being hidden by the early anomalous behavior).

\paragraph{GRB 970815, GRB 030131}
In each case, the afterglow light curve has only two data points, and a late
upper limit indicates a faint host. As the degree of freedom
is zero, no errors are given.

\paragraph{GRB 980326}
The light curve fit includes a supernova component with $k=1$, $s=1$ fixed
(equation~\ref{mag1}), assuming a redshift of $z=1$ \citep[][]{BloomNat}.

\paragraph{GRB 980519}
The first data point of the $R_C$ band light curve of  the afterglow of GRB
980519 is at $t\approx0.5$ days after the burst. In the $I_C$ band and $V$
band earlier data points exist but no late-time data. In order to improve the
fit, we assumed achromacy and mixed these bands by shifting the $I_C$ and $V$
band to the same zero point as the $R_C$ band, fitting the composite light
curve. For $t_b/(1+z)$, we assumed a redshift of $z=1.5$, since
\cite{Jaunsen2001} state that $z\ge1.5$ from the absence of a supernova bump
in the light curve. Using $z=1.5$ also gives a very good SED fit (Paper III).

\paragraph{GRB 990123 and GRB 021211}
These afterglow light curves have very early detections, where the light curve
is dominated by reverse shock emission. Data from the reverse shock dominated
phase have not been included in the fits. For GRB 021211 the result are
sensitive to the data used for the fit. We used only data with $t>0.014$ days.
See also our comment in \S~\ref{data}.

\paragraph{GRB 990705, GRB 020322, GRB 020410, GRB 030115, GRB 030324,
GRB 031220, GRB 040422}
In all these cases, afterglow data are too sparse to confine certain
parameters of the light curves. The addition of observational upper limits
was however used in order to derive  upper or lower limits on these
parameters. For light curve decay slopes, the sequence data point - upper
limit - host magnitude sets a lower limit on the decay rate (e.g., GRB 020410)
as the observational upper limit does not preclude an even steeper decay rate.
On the other hand, the sequence upper limit - data point - host magnitude sets
an upper limit on the decay rate (e.g., GRB 030324), as the upper limit cannot
preclude a slower decay rate. In the case of GRB 040422, early unfiltered
upper limits were, after correction for Galactic extinction, brightened by
three magnitudes (the approximate typical $R_C-K$ color) to get early $K$-band
upper limits. In some cases early data was adequate to derive $\alpha_1$ and a
later upper limit lies beneath the extrapolated light curve decay, indicating
a break in the light curve must have occurred.

\paragraph{GRB 991208, GRB 000131, GRB 001007}
In all three cases, the optical afterglows were located several days after the
burst and exhibited a steep decay $\alpha\gr2$. It is highly probable that the
first observations are after the jet break, meaning that the decay slope is
$\alpha_2$. For GRB 991208, the upper limit on $\alpha_1$ stems from a very
shallow but very early upper limit \citep{Castro-Tirado2001}. We note that a
postbreak decay slope is also possible for GRB 000911 \citep{Masetti2005} --
if so, it would be quite shallow. There are also several GRBs (e.g., GRB
990308, GRB 001011) where data quality is so sparse that no evidence for or
against a break in the afterglow light curve can be found.

\paragraph{GRB 011121, GRB 020405}
These bursts have relatively bright or structured hosts, which makes it
difficult to extract the afterglow light curve. Different groups used
different methods to do this. Therefore we decided not to mix the late data
points (where the different host galaxy subtraction methods lead to
significantly different magnitudes), but instead to use the data set from only
one group in each case. The late data set of GRB 011121 is extracted from
\cite{Greiner2003}, the data set from GRB 020405 is provided by N. Masetti
(private communication).

\paragraph{GRB 020305, GRB 030725}
The late-time rebrightenings of these afterglow light curves, which have been
attributed to possible supernova components \citep{Gorosabel2005,
Pugliese2005}, have not been included in the fit as no reliable redshift
estimate is known. In the case of GRB 030725, there is a large data gap from
0.5 to 4.5 days. A fit that leaves $t_b$ as a free parameter is not confined
in $\alpha_1$ and $t_b$, thus, $t_b$ was fixed at a reasonable value.

\paragraph{GRB 021004, GRB 030329}
The light curves of these afterglows are clear outliers compared to the rest
of our sample and very complicated to fit. The light curves each show several
re-brightening episodes, which cannot be fit correctly with a smoothly broken
power-law. As especially the light curve of the afterglow of GRB 030329 has
been analyzed in detail \citepeg{Lipkin2004}, we did not do this again. As
with the other afterglows, we fit smoothly broken power-laws to the light
curves (the results are given in Table \ref{tabres}). In the case of GRB
021004, the anomalous behavior at $t<0.07$ days is excluded from the fit. We
performed these fits for completeness, even though the $\chi^2_{\rm d.o.f.}$
values  ($\chi^2_{\rm d.o.f.} >20$ for both) show that they are not well
approximated by equation (\ref{mag1}). While $\alpha_1$ is almost
unaffected, $\alpha_2$ and $t_b$ are highly dependent upon the value of the
smoothness parameter $n$, which had to be fixed. These results are thus not
included in our statistical analysis.

If we concentrate on certain parts of the light curve of the afterglow of  GRB
030329, they can be very well fit by a smoothly broken power-law. We have
performed two additional fits, one using only data up to 0.55 days, before the
time of the probable jet break \citepeg{Uemura2003}, and thus  encompassing
the cooling break \citep{Sato2003}, the other one using data from 0.28 days up
to 1 day after the burst (the beginning of the first rebrightening) and thus
encompassing the supposed  jet break. In the former case, $n$ is not confined,
the fit formally finds $n=450\pm2800$, but a very sharp transition is indicated,
and thus we fix $n=100$. In the last case, we were able to let
the smoothness parameter $n$ vary and obtain  $n=7.54\pm1.47$ for the jet
break. The values we derive for the jet break, $\alpha_1=1.17\pm0.01$ and
$\alpha_2=2.21\pm0.07$, are both unremarkable and lie well within the
distribution we find for the 16 afterglows we study here (cf. Figure
\ref{a1Histo}). The slope change $\Delta\alpha$ is $\Delta\alpha=0.33\pm0.01$
for the cooling break \citep[slightly higher than the theoretical prediction of
$\Delta\alpha=0.25$,][]{PK2001a} and $\Delta\alpha=1.04\pm0.10$ for the jet break.
This value is also unremarkable (cf. Figure \ref{DaHisto}). The rest frame break
time is $0.58\pm0.01$ days after the burst, once again typical for our sample
of 16 afterglows  (cf. Figure \ref{logtbz}). The values of $\alpha_1$ and
$\alpha_2$ lie very well on the theoretical line describing a Wind/ISM model
with $\nu > \nu_c$. It is one of only a few bursts that are found in this
region (cf. $\S$ \ref{profile}, Figure \ref{a1a2}). As the cooling frequency
has passed the optical bands before the jet break and now lies at longer
wavelengths, this implies an evolution from high to low frequencies and thus
a wind model \citep{Chevalier2000}.

\paragraph{GRB 030323}
The afterglow of this GRB has a late break which is only represented by one
data point. $\alpha_2$ was fixed to a value derived from a free fit to derive
meaningful errors on $\alpha_1$ and $t_b$.

\paragraph{XRF 030723}
The early light curve of this afterglow has a plateau phase. We do not fit
this plateau phase but start at 0.9 days after the burst. While a light curve
break is evident, the prebreak data are inadequate to give more than
limits. We also do not fit the late time data points ($t>8$ days), which show
a significant rebrightening \citep[][]{Fynbo2004} which can be modelled with a
SN light curve only when it rises much steeper than SN1998bw. Another
explanation could be that this rebrightening is caused by a two component jet
\citep{Huang2004}.

\paragraph{GRB 040827}
The steep decay of this light curve ($\alpha\ge2$) indicates that this is a
postbreak decay, but the data quality is low and the error large enough to
make this conclusion unsure.

\paragraph{GRB 040924}
The light curve is fit with an unbroken power-law, excluding the earliest
data point \citep{Fox2005} from the fit.

\paragraph{GRB 041006}
As the photometric calibration of the very earliest data point of the optical
afterglow \citep{MaenoGCN} is unsure, it is not included in the fit. Including
it would strongly reduce $\alpha_1$ and $n$.

\section{References of host magnitudes \label{hosts}}

The following magnitudes have not been corrected for Galactic extinction and
thus differ from those displayed in Table~\ref{tabres}.

\begin{itemize}
\itemsep-1mm
\item GRB 990308: $R_C=29.4\pm0.4$ mag, \cite{Jaunsen2003}
\item GRB 001011: $R_C=25.38\pm0.25$ mag, \cite{Gorosabel2002}
\item GRB 020305: $R_C=25.17\pm0.14$ mag, \cite{Gorosabel2005}
\item GRB 021004: $R_C=24.21\pm0.04$ mag, \cite{deUga05}
\item GRB 030115: $R_C=25.2\pm0.3$ mag, \cite{Dullighan2004}
\item GRB 030324: $R_C=25.16\pm0.24$ mag, \cite{NysewanderGCN}
\item GRB 030329: $R_C=22.66\pm0.04$ mag, \cite{GorosabelACC}
\end{itemize}



\begin{rotate}

\begin{deluxetable}{l r c c c c c c c c c c}
\tablecolumns{12}
\tablewidth{0pc}
\tablecaption{Results of the light curve fitting (Equation~\ref{mag1})}
\tablehead{
\colhead{GRB}   &
\colhead{\#} &
\colhead{band} &
\colhead{$\chi^2_{dof}$} &
\colhead{$d.o.f.$} &
\colhead{$m_c$}    &
\colhead{$\alpha_1$}  &
\colhead{$\alpha_2$}  &
\colhead{$t_b$}    &
\colhead{$n$}      &
\colhead{$m_{\rm host}$\tablenotemark{a}}}
\startdata
970228  &    & $R_C$    & 0.70        & 4    & 20.43$\pm$0.21   & 1.46$\pm$0.15   & \nodata         & \nodata       & \nodata       & 24.65$\pm$0.51    \\
970508  &    & $R_C$    & 3.95        & 61   & 18.64$\pm$0.02   & 1.24$\pm$0.01   & \nodata         & \nodata       & \nodata       & 25.29$\pm$0.09    \\
970815  &    & $I_C$    & \nodata     & 0    & 21.73            & 0.34            & \nodata         & \nodata       & \nodata       & 25                \\
971214  &    & $R_C$    & 1.12        & 10   & 22.91$\pm$0.05   & 1.49$\pm$0.08   & \nodata         & \nodata       & \nodata       & 25.64$\pm$0.05    \\
980326  &    & $R_C$    & 2.74        & 16   & 22.82$\pm$0.04   & 1.85$\pm$0.05   & \nodata         & \nodata       & \nodata       & 28.95$\pm$0.53    \\
980329  &    & $R_C$    & 2.85        & 8    & 23.91$\pm$0.17   & 0.85$\pm$0.12   & \nodata         & \nodata       & \nodata       & 26.67$\pm$0.10    \\
980519  &  1 & $R_C$\tablenotemark{b} & 2.73 & 68 & 18.86$\pm$0.13 & 1.50$\pm$0.12 & 2.27$\pm$0.03  & 0.48$\pm$0.03 & 10            & 25.36$\pm$0.12    \\
980613  &    & $R_C$    & $<0.01$     & 1    & 23.07$\pm$0.21   & 0.44$\pm$0.23   & \nodata         & \nodata       & \nodata       & 24.04$\pm$0.50    \\
980703  &    & $R_C$    & 0.77        & 13   & 21.51$\pm$0.96   & 0.85$\pm$0.84   & 1.65$\pm$0.46   & 1.35$\pm$0.94 & 10            & 22.46$\pm$0.08    \\
990123  &  2 & $R_C$    & 2.11        & 44   & 21.37$\pm$0.60   & 1.24$\pm$0.06   & 1.62$\pm$0.15   & 2.06$\pm$0.83 & 10            & 23.99$\pm$0.09    \\
990308  &    & $R_C$    & $<0.01$     & 1    & 22.28$\pm$4.00   & \nodata         & 1.96$\pm$1.89   & \nodata       & \nodata       & 29.34             \\
990510  &  3 & $V$      & 1.57        & 59   & 19.50$\pm$0.05   & 0.92$\pm$0.02   & 2.10$\pm$0.06   & 1.31$\pm$0.07 & 2.25$\pm$0.51 & 28.37$\pm$0.48    \\
990705  &    & $H$      & \nodata     & \nodata  & $\ge18.3$    & $\le1.68$       & $\ge2.88$       & $\ge0.75$     & 10            & 22                \\
990712  &    & $R_C$    & 1.27        & 18   & 21.22$\pm$0.02   & 0.96$\pm$0.02   & \nodata         & \nodata       & \nodata       & 21.80$\pm$0.02    \\
991208  &    & $R_C$    & 1.74        & 12   & 16.60$\pm$0.07   & $\le1.38$       & 2.47$\pm$0.05   & $\le2.08$     & \nodata       & 24.28$\pm$0.16    \\
991216  &  4 & $R_C$    & 1.47        & 65   & 18.09$\pm$0.18   & 1.17$\pm$0.03   & 1.57$\pm$0.03   & 1.10$\pm$0.13 & 10            & 23.52$\pm$0.09    \\
000131  &    & $R_C$    & 0.18        & 1    & 19.88$\pm$0.31   & \nodata         & 2.40$\pm$0.21   & \nodata       & \nodata       & 27                \\
000301  &  5 & $R_C$    & 4.93        & 50   & 20.70$\pm$0.06   & 0.57$\pm$0.05   & 2.81$\pm$0.13   & 4.93$\pm$0.18 & 2.36$\pm$0.67 & 27.95$\pm$0.30    \\
000418  &    & $R_C$    & 1.68        & 16   & 23.18$\pm$0.94   & 1.15$\pm$0.41   & 2.69$\pm$0.66   & 7.85$\pm$2.71 & 10            & 23.46$\pm$0.03    \\
000630  &    & $R_C$    & 0.46        & 5    & 23.19$\pm$0.06   & 1.12$\pm$0.11   & \nodata         & \nodata       & \nodata       & 26.68$\pm$0.21    \\
000911  &    & $R_C$    & 0.34        & 8    & 19.67$\pm$0.09   & 1.46$\pm$0.04   & \nodata         & \nodata       & \nodata       & 25.11$\pm$0.11    \\
000926  &  6 & $R_C$    & 1.10        & 49   & 20.81$\pm$0.16   & 1.74$\pm$0.03   & 2.45$\pm$0.05   & 2.10$\pm$0.15 & 10            & 25.22$\pm$0.06    \\
001007  &    & $R_C$    & 0.52        & 4    & 17.48$\pm$0.22   & \nodata         & 2.06$\pm$0.13   & \nodata       & \nodata       & 24.73$\pm$0.15    \\
001011  &    & $R_C$    & $<0.01$     & 1    & 22.45$\pm$0.16   & 1.45$\pm$0.14   & \nodata         & \nodata       & \nodata       & 25.19             \\
010222  &  7 & $R_C$    & 2.05        & 133  & 19.21$\pm$0.24   & 0.60$\pm$0.09   & 1.44$\pm$0.02   & 0.64$\pm$0.09 & 2.29$\pm$0.68 & 26.68$\pm$0.17    \\
010921  &    & $r'$     & 0.97        & 4    & 19.46$\pm$0.03   & 1.56$\pm$0.07   & \nodata         & \nodata       & \nodata       & 21.63$\pm$0.02    \\
011121  &  8 & $R_C$    & 1.27        & 17   & 20.27$\pm$0.32   & 1.76$\pm$0.05   & 2.99$\pm$0.28   & 1.54$\pm$0.22 & 10            & host corrected    \\
011211  &  9 & $R_C$    & 7.21        & 43   & 21.72$\pm$0.15   & 0.93$\pm$0.02   & 2.31$\pm$0.27   & 2.34$\pm$0.34 & 10            & host corrected    \\
020124  & 10 & $R_C$    & 0.71        & 10   & 22.85$\pm$1.00   & 1.47$\pm$0.06   & 2.12$\pm$0.27   & 1.36$\pm$0.77 & 10            & 30.68$\pm$2.28    \\
020305  &    & $R_C$    & 3.38        & 4    & 19.60$\pm$0.20   & 1.19$\pm$0.07   & \nodata         & \nodata       & \nodata       & 25.04             \\
020322  &    & $R_C$    & $<0.01$     & 1    & 23.66$\pm$0.49   & 0.45$\pm$0.39   & $\ge$2.17       & 0.95$\pm$0.27 & 10            & host corrected    \\
020331  &    & $R_C$    & 1.98        & 6    & 22.56$\pm$0.26   & 0.69$\pm$0.04   & 2.12$\pm$0.40   & 7.17$\pm$1.52 & 10            & 24.89$\pm$0.16    \\
020405  & 11 & $R_C$    & 5.26        & 12   & 21.35$\pm$0.32   & 1.26$\pm$0.09   & 1.93$\pm$0.13   & 2.40$\pm$0.45 & 10            & host corrected    \\
020410  &    & $R_C$    & \nodata     & \nodata & $\le22.4$     & $\ge1.25$       & \nodata         & \nodata       & \nodata       & 28.23$\pm$0.5     \\
020813  & 12 & $R_C$    & 2.00        & 59   & 19.27$\pm$0.11   & 0.67$\pm$0.07   & 1.78$\pm$0.28   & 0.77$\pm$0.25 & 1.44$\pm$1.06 & 23.61$\pm$0.15    \\
020903  &    & $R_C$    & 1.52        & 3    & 19.54$\pm$0.21   & 1.27$\pm$0.58   & \nodata         & \nodata       & \nodata       & 20.91$\pm$0.47    \\
021004  &    & $R_C$    & 38.5        & 378  & 21.62$\pm$0.02   & 1.07$\pm$0.01   & 2.12$\pm$0.07   & 8.62$\pm$0.16 & 10            & 24.06             \\
021211  &    & $R_C$    & 2.00        & 27   & 20.30$\pm$0.90   & 0.96$\pm$0.04   & 1.22$\pm$0.10   & 0.11$\pm$0.09 & 10            & 25.20$\pm$0.12    \\
030115  &    & $R_C$    & 0.10        & 1    & $\ge20.30$       & 0.44$\pm$0.12   & $\ge3$          & $\ge2$        & 10            & 24.8              \\
030131  &    & $R_C$    & \nodata     & 0    & 23.35            & 1.06            & \nodata         & \nodata       & \nodata       & 30                \\
030226  & 13 & $R_C$    & 3.86        & 35   & 19.67$\pm$0.33   & 0.58$\pm$0.16   & 2.68$\pm$0.28   & 0.96$\pm$0.10 & 0.91$\pm$0.49 & 27.1              \\
030227  &    & $R_C$    & 1.07        & 4    & 22.83$\pm$0.11   & 1.18$\pm$0.15   & \nodata         & \nodata       & \nodata       & 25                \\
030323  &    & $R_C$    & 2.16        & 36   & 22.94$\pm$0.18   & 1.36$\pm$0.02   & 2.7             & 6.71$\pm$0.74 & 10            & 27.86$\pm$0.52    \\
030324  &    & $I_C$    & \nodata     & \nodata    & $\le25.65$ & $\le1.32$       & \nodata         & \nodata       & \nodata       & 25                \\
030328  & 14 & $R_C$    & 1.34        & 18   & 20.61$\pm$0.23   & 0.87$\pm$0.04   & 1.54$\pm$0.11   & 0.60$\pm$0.10 & 10            & 24.15$\pm$0.35    \\
030329  &    & $R_C$    & 30.4        & 2953 & 17.63$\pm$0.01   & 1.10$\pm$0.01   & 2.32$\pm$0.01   & 5.27$\pm$0.02 & 10            & 22.60             \\
030329\tablenotemark{c}  &    & $R_C$    & 0.85        & 1165 & 13.92$\pm$0.01   & 0.86$\pm$0.01   & 1.19$\pm$0.01   & 0.27$\pm$0.01 & 100 & 22.60             \\
030329\tablenotemark{d}  &    & $R_C$    & 0.64        & 946  & 15.11$\pm$0.03   & 1.17$\pm$0.01   & 2.21$\pm$0.07   & 0.68$\pm$0.02 & 7.54$\pm$1.47 & 22.60             \\
030418  &    & $R_C$    & 0.42        & 10   & 22.22$\pm$1.31   & 1.23$\pm$0.09   & 1.72$\pm$0.48   & 1.50$\pm$1.26 & 10            & 27                \\
030429  & 15 & $R_C$    & 7.68        & 11   & 21.80$\pm$0.08   & 0.81$\pm$0.03   & 3.03$\pm$0.27   & 2.17$\pm$0.09 & 10            & 27                \\
030528  &    & $Ks$     & 0.53        & 1    & 19.28$\pm$0.65   & 0.73$\pm$0.89   & \nodata         & \nodata       & \nodata       & 19.82$\pm$0.75    \\
030723  &    & $R_C$    & 1.61        & 12   & $\le21.45$       & $\le0.88$       & 2.12$\pm$0.06   & $\le1.57$     & 10            & 27                \\
030725  &    & $R_C$    & 1.31        & 8    & 20.45$\pm$0.05   & 0.80$\pm$0.06   & 1.65$\pm$0.06   & 2.9           & 10            & 25                \\
031203  &    & $J$      & 0.20        & 24   & 19.36$\pm$0.98   & 0.69$\pm$0.50   & \nodata         & \nodata       & \nodata       & 17.43$\pm$0.15    \\
031220  &    & $R_C$    & \nodata     & \nodata    & $\le23.7$  & $\le1.77$       & \nodata         & \nodata       & \nodata       & 23.13$\pm$0.11    \\
040106  &    & $R_C$    & 0.05        & 1    & 22.86$\pm$0.10   & 1.31$\pm$0.11   & \nodata         & \nodata       & \nodata       & 28                \\
040422  &    & $Ks$     & \nodata     & \nodata  & $\le21.28$   & $\le1.3$        & \nodata         & \nodata       & \nodata       & 19.74$\pm$0.17    \\
040827  &    & $Ks$     & 1.52        & 11   & 21.05$\pm$0.34   & \nodata         & 2.08$\pm$0.45   & \nodata       & \nodata       & 20.00$\pm$0.05    \\
040916  &    & $R_C$    & 0.59        & 3    & 23.64$\pm$0.11   & 0.96$\pm$0.07   & \nodata         & \nodata       & \nodata       & 30                \\
040924  &    & $R_C$    & 1.37        & 29   & 22.96$\pm$0.04   & 1.09$\pm$0.02   & \nodata         & \nodata       & \nodata       & 24.55$\pm$0.19    \\
041006  & 16 & $R_C$    & 1.25        & 81   & 19.45$\pm$0.27   & 0.68$\pm$0.06   & 1.30$\pm$0.02   & 0.23$\pm$0.04 & 4.87$\pm$2.57 & 28.4              \\ \hline
\enddata
\tablenotetext{a}{ If a host magnitude is given with an error then this is the
result of the fit. Otherwise we had to fix this value because the data set at
late times is too sparse. In such cases we either used the host magnitudes
reported in the literature (GRBs 990308, 001011, 020305, 021004, 030115,
030324, 030329; see \S~\ref{hosts}) or we used a reasonable estimate (GRBs
970815, 990705, 000131, 030131, 030226, 0303227, 030418, 030429, 030723,
030725, 040106, 040916, 041006).}
\tablenotetext{b}{See \S~\ref{individual}.}
\tablenotetext{c}{This fit uses only data up to 0.55 days after the burst, and
encompasses the probable cooling break \citep{Sato2003}.
See \S~\ref{individual}.}
\tablenotetext{d}{This fit uses only data from 0.28
days to 1 day after the burst, and encompasses the supposed jet break
\citepeg{Uemura2003}. See \S~\ref{individual}.}
\label{tabres}
\end{deluxetable}
\end{rotate}


\clearpage
\thispagestyle{empty}
\setcounter{page}{24}

\begin{rotate}

\begin{deluxetable}{crrrrrrrrrr}
\tablecolumns{12}
\tablewidth{0pc}
\tablecaption{The absolute value of the correlation coefficient between the various parameters}
\tablehead{
\colhead{}
& \colhead{$\Delta\alpha$}
& \colhead{$t_b/(1+z)$}
& \colhead{$\theta_{\rm jet, ISM}$}
& \colhead{$\theta_{\rm jet, wind}$}
& \colhead{$\theta_{\rm jet, mixed}$\tablenotemark{a}}
& \colhead{$\log(E_{\rm cor, ISM})$}
& \colhead{$\log(E_{\rm cor, wind})$}
& \colhead{$\log(E_{\rm cor, mixed})$\tablenotemark{a}}
& \colhead{$\log(E_{\rm peak})$}
& \colhead{$z$} }
\startdata
$\alpha_1$                & 0.50         & 0.30        & 0.55        & 0.33         & 0.02         & 0.29          & 0.39           & 0.37          & 0.46      & 0.20 \\
$\alpha_2$                & 0.65         & 0.46        & 0.69        & 0.74         & 0.66         & 0.49          & 0.51           & 0.50          & 0.77      & 0.75 \\
$\Delta\alpha$            & \nodata      & 0.62        & 0.36        & 0.55         & 0.37         & 0.40          & 0.19           & 0.24          & 0.31      & 0.55 \\
$t_b/(1+z)$               & \nodata      & \nodata     & 0.82        & 0.58         & 0.62         & 0.28          & 0.31           & 0.29          & 0.01      & 0.53 \\
$\theta_{\rm jet, ISM}$   & \nodata      & \nodata     & \nodata     & \nodata      & \nodata      & 0.14          & 0.37           & 0.28          & 0.18      & 0.00 \\
$\theta_{\rm jet, wind}$  & \nodata      & \nodata     & \nodata     & \nodata      & \nodata      & 0.49          & 0.66           & 0.67          & 0.53      & 0.14 \\
$\theta_{\rm jet, mixed}$\tablenotemark{a}   & \nodata      & \nodata     & \nodata     & \nodata      & \nodata      & 0.43          & 0.58           & 0.52          & 0.55      & 0.15 \\
$\log(E_{\rm cor, ISM})$  & \nodata      & \nodata     & \nodata     & \nodata      & \nodata      & \nodata       & \nodata        & \nodata       & 0.81      & 0.05 \\
$\log(E_{\rm cor, wind})$ & \nodata      & \nodata     & \nodata     & \nodata      & \nodata      & \nodata       & \nodata        & \nodata       & 0.91      & 0.03 \\
$\log(E_{\rm cor, mixed})$\tablenotemark{a}  & \nodata      & \nodata     & \nodata     & \nodata      & \nodata      & \nodata       & \nodata        & \nodata       & 0.78      & 0.03 \\
$\log(E_{\rm peak})$      & \nodata      & \nodata     & \nodata     & \nodata      & \nodata      & \nodata       & \nodata        & \nodata       & \nodata   & 0.01 \\ \hline
\enddata
\tablenotetext{a}{Here the data are calculated according to an ISM or a wind
model depending on the location of the corresponding burst in the
$\alpha_1-\alpha_2$-diagram (Fig.~\ref{a1a2} and \S~\ref{theta}).}
\label{tabcor}
\end{deluxetable}

\end{rotate}


\clearpage
\pagestyle{plaintop}

\begin{figure*}[htp]
\includegraphics[width=.8\textwidth]{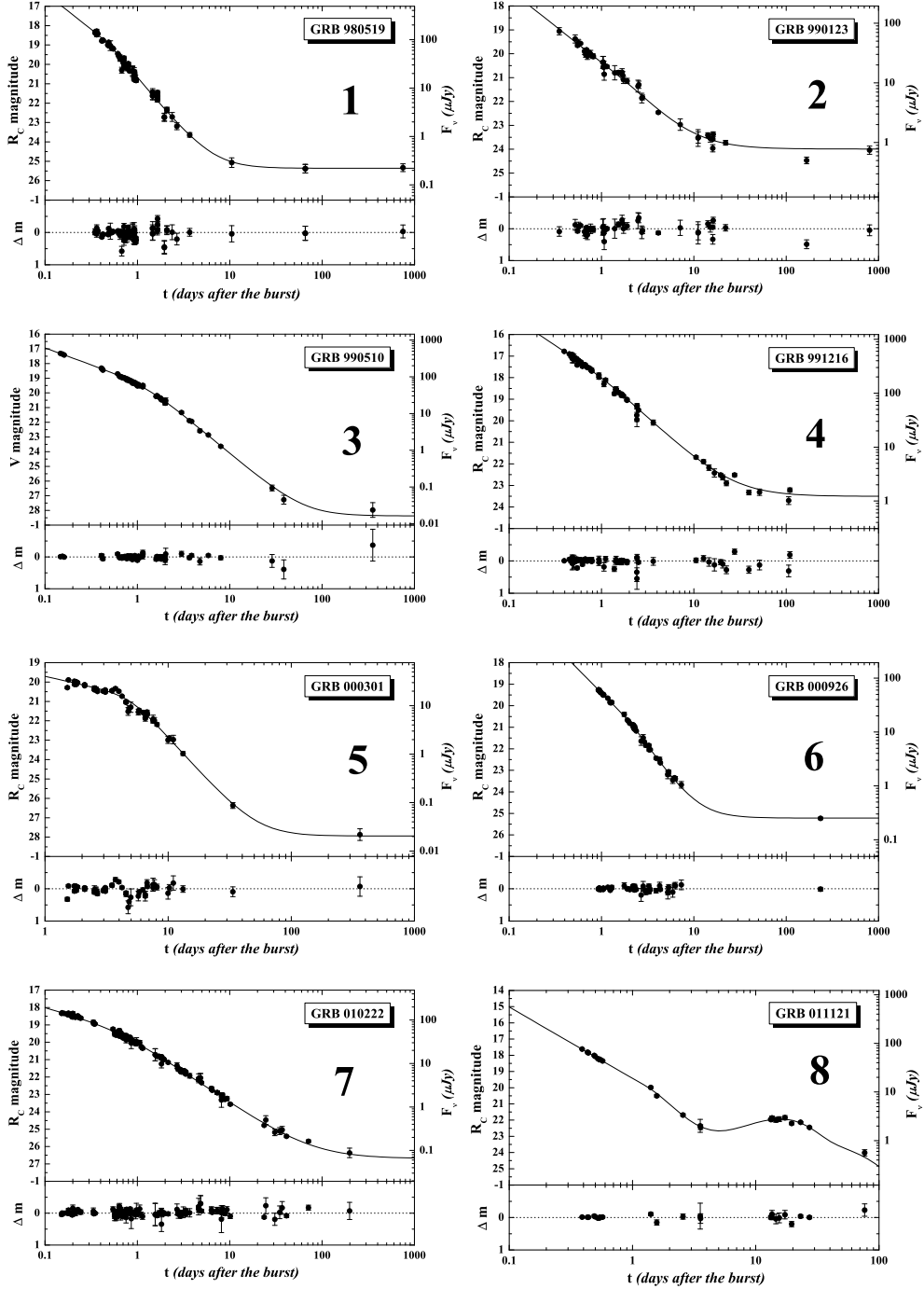}
\caption[]{The light curves of the 16 afterglows we have investigated.
In all but one case (GRB 990510) we have analyzed the $R_C$-band data.
$\Delta m$ is the difference between the observed and the fitted magnitude.
For GRBs 011121, 011211, and 020405 the data are corrected for
the flux from the underlying host galaxy. For GRBs 030226 and 041006
the $R_C$-band magnitude of the host was assumed to be 27.1 and 28.4,
respectively. The numbering follows Table~\ref{tabres}.}
\label{GS1}
\end{figure*}

\newpage\clearpage

\begin{figure*}[htp]
\includegraphics[width=.8\textwidth]{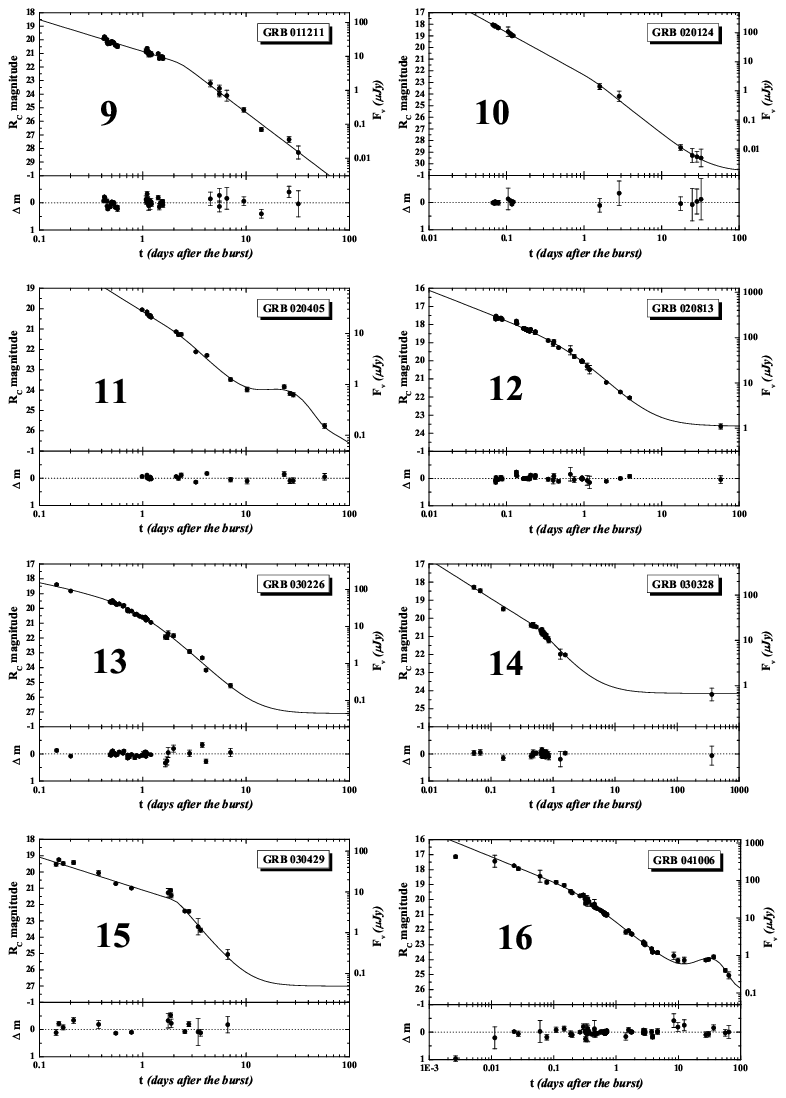}
\addtocounter{figure}{-1}
\caption[]{continued.}
\label{GS2}
\end{figure*}

\newpage\clearpage

\begin{figure}[htp]
\includegraphics[width=1\textwidth]{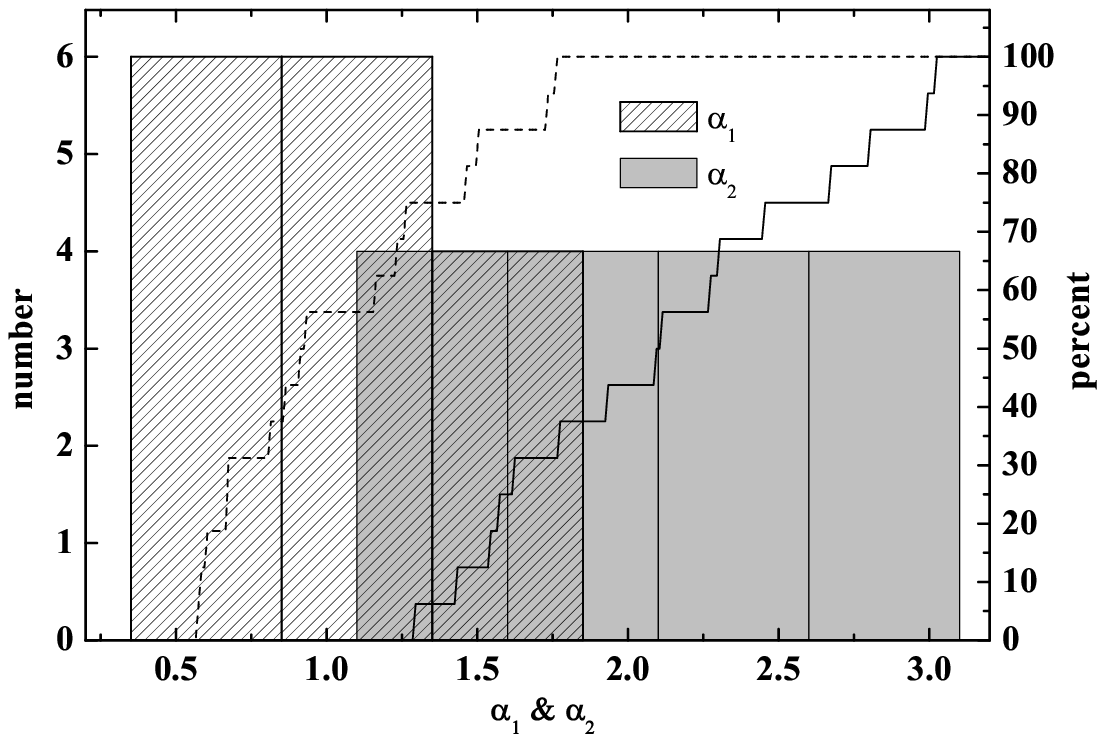}
\caption[]{
The distribution of the prebreak decay slope $\alpha_1$ and the postbreak
decay slope $\alpha_2$ for the 16 afterglows in our sample with the best
defined light curves (Table~\ref{tabres}; bin size 0.5). The broken line
and the solid line represent the cumulative distribution for $\alpha_1$
and $\alpha_2$, respectively.}
\label{a1Histo}
\end{figure}

\newpage\clearpage

\begin{figure}[htp]
\includegraphics[width=1\textwidth]{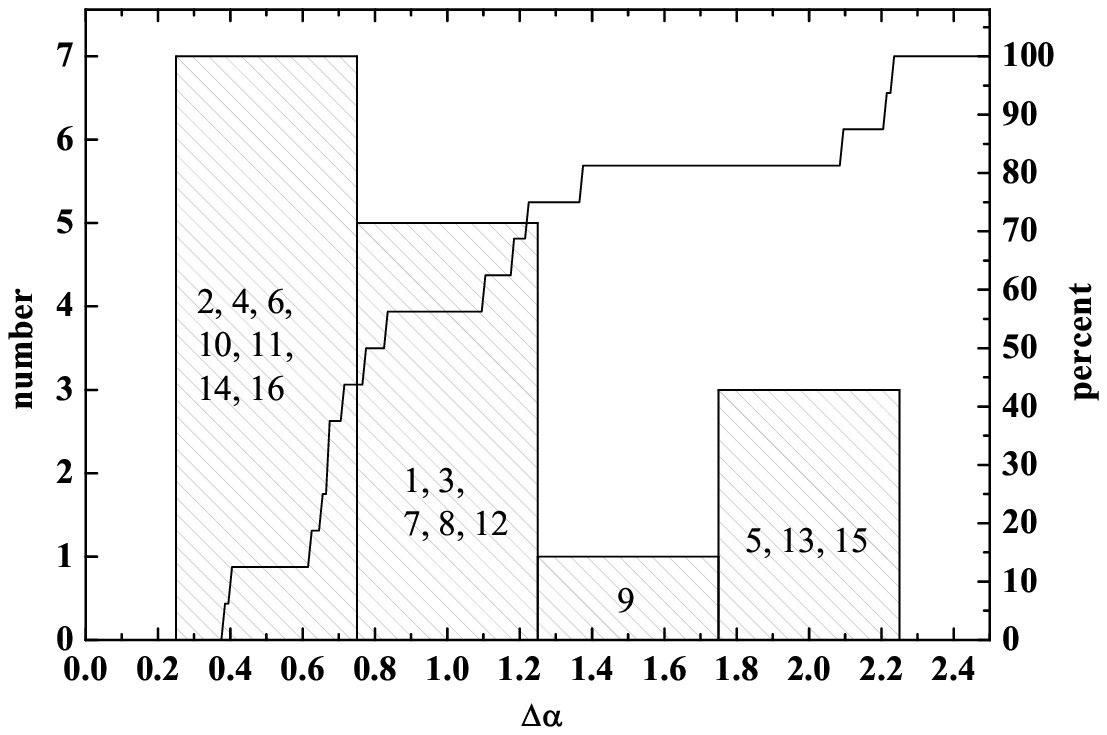}
\caption[]{
The distribution of the light curve steepening  $\Delta \alpha =
\alpha_2-\alpha_1$. The numbering follows Table~\ref{tabres}. The  solid line
is the cumulative distribution.}
\label{DaHisto}
\end{figure}

\newpage\clearpage

\begin{figure}[htp]
\includegraphics[width=1\textwidth]{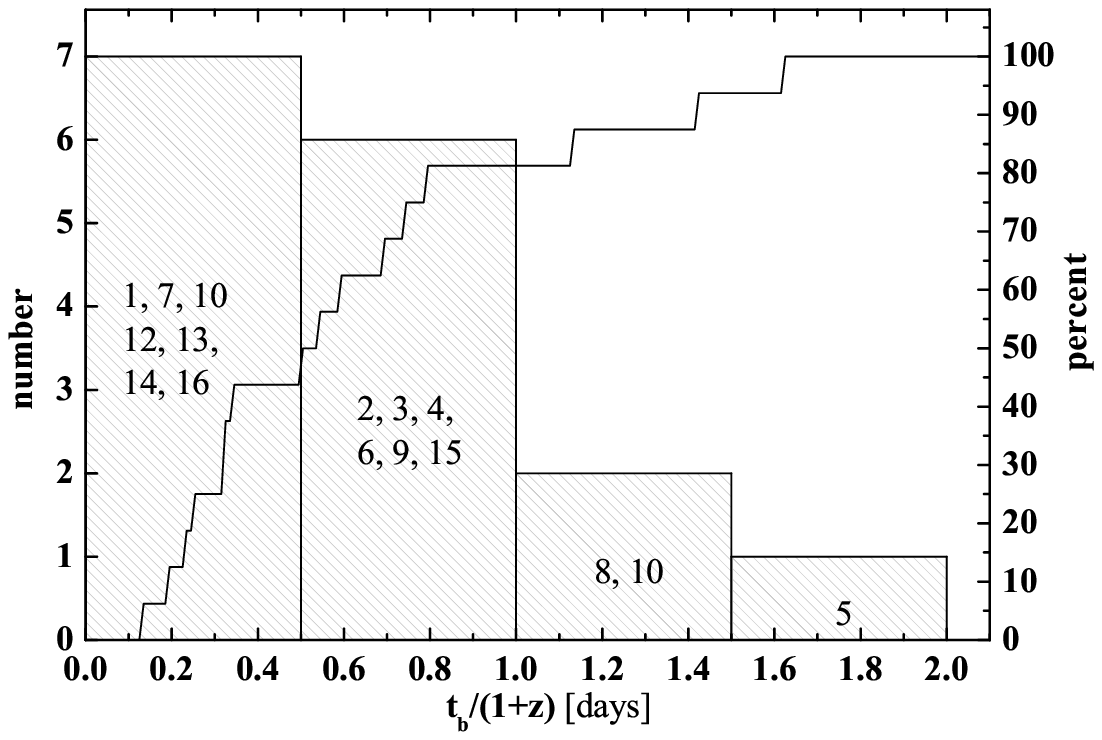}
\caption[]{
The distribution of the break time $t_{\rm b}$ in the GRB host frame for all
bursts of our sample of 16 afterglows. The data cover the range
from $t_{\rm b}/(1+z)=0.14\pm0.02$ days (GRB 041006) to $1.62\pm0.06$ days
(GRB 000301C). Most afterglows exhibit a break at less than  1 day after the
burst in the host frame. For GRB 980519 we assumed a redshift of $z=1.5$, but
it would fall into the first bin even if $z=0$. The solid line represents
the cumulative distribution.}
\label{logtbz}
\end{figure}

\newpage\clearpage

\begin{figure}[htp]
\includegraphics[width=1\textwidth]{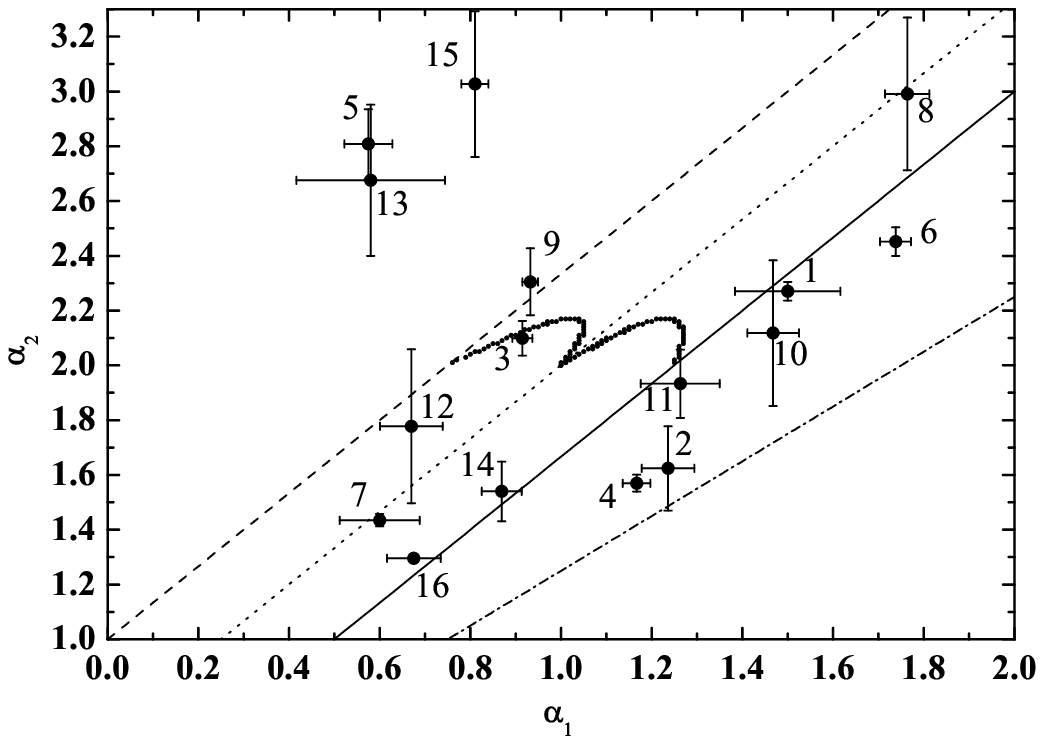}
\caption[]{
The afterglow parameters $\alpha_1$ and $\alpha_2$ in comparison with standard
afterglow models \citep{PK2001a}. The dashed line is for an ISM density
profile with $\nu < \nu_c$, the dotted line for the ISM/wind model with $\nu >
\nu_c$ and a Compton parameter, $Y$, less than 1, and the solid line for the
wind model with $\nu < \nu_c$. Note that we have extended these curves to
$\alpha_2 < 2.0$. The two curves around ($\alpha_1=1.0, \alpha_2=2.0$)
represent the case $\nu > \nu_c$, $Y>1, 2<p<3$ \citep{PK2001a}. The
dash-dotted line stands for the theoretical prediction of the  passage of the
cooling break ($\alpha_2 = \alpha_1+0.25$). While GRB 990123 (\# 2) lies on
this line within errors, it has a late achromatic break which is very probably
a shallow jet break.}
\label{a1a2}
\end{figure}

\newpage\clearpage

\begin{figure}[htp]
\includegraphics[width=1\textwidth]{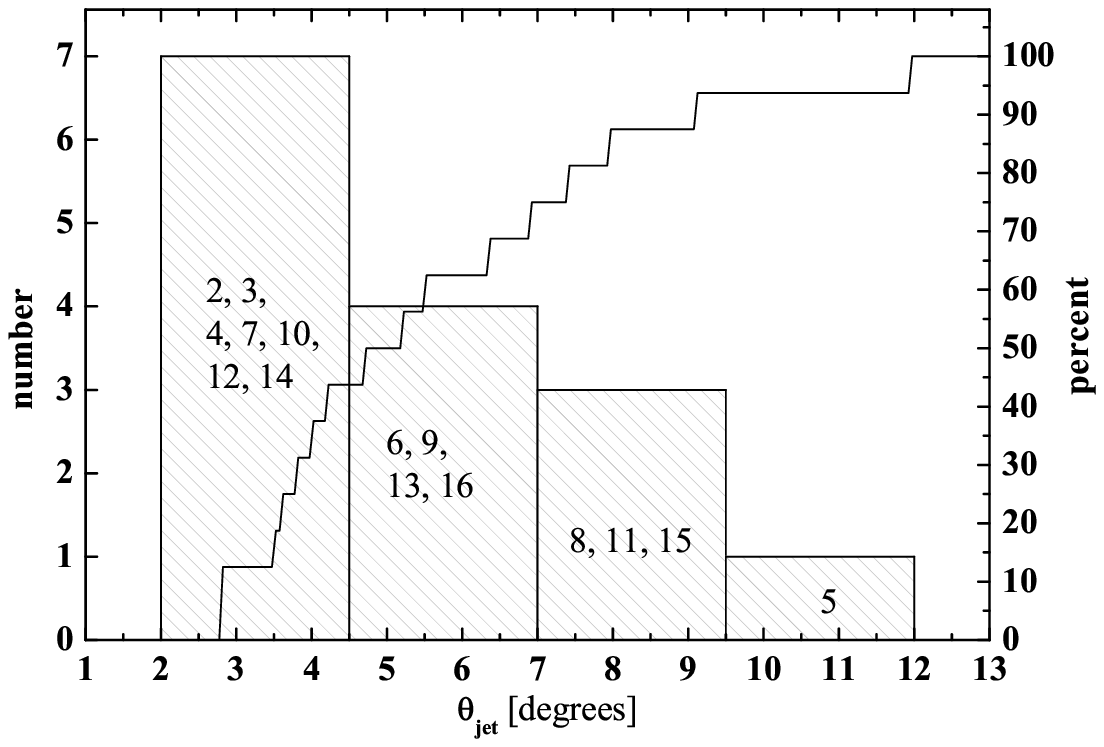}
\caption[]{
The distribution of the derived jet half-opening angle, $\theta_{\rm jet}$, of
our sample (\S~\ref{theta}). GRB 980519 is not included here since its
redshift is not exactly known.}
\label{thetaHisto}
\end{figure}

\newpage\clearpage

\begin{figure}[htp]
\includegraphics[width=1\textwidth]{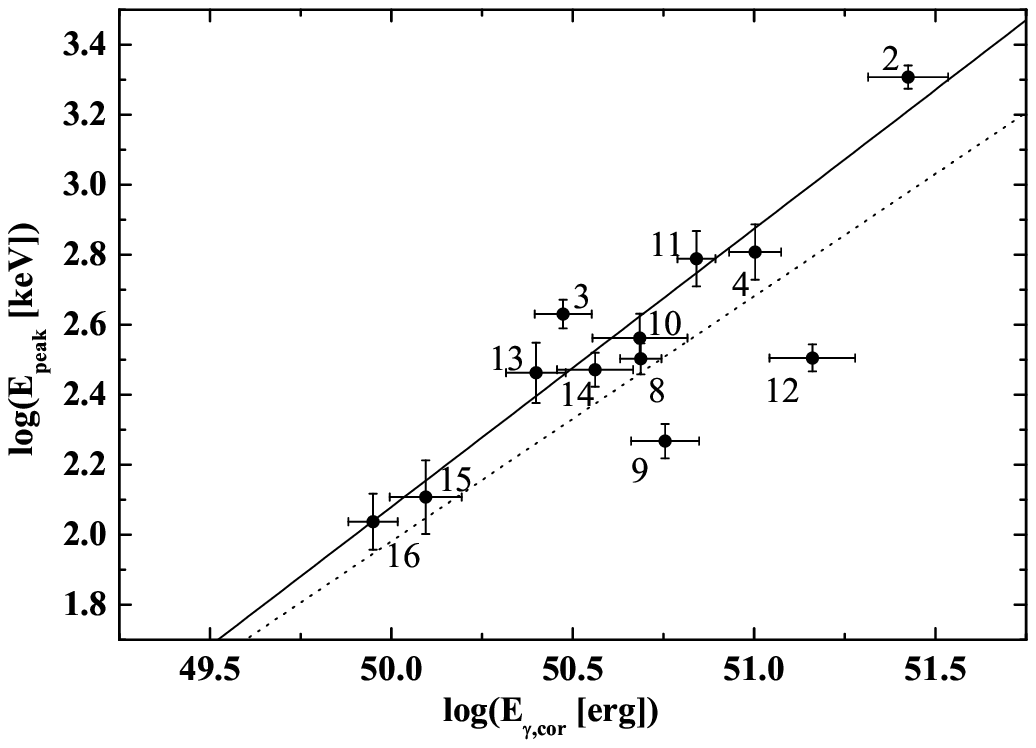}
\caption[]{
The Ghirlanda relation (Ghirlanda et al. 2004) as it follows from our light
curve data (Table~\ref{tabres}, \S~\ref{theta}) in combination with the
high-energy data given in \cite{Friedman2005}.  GRB \#1 is not shown because
of the uncertainty of its redshift and GRBs \#5, \#6, \#7 are not included
here since $E_{\rm peak}$ is not known.  Our fit gives a slope of
$0.79\pm0.09$ if the outliers GRB 011211 (\# 9) and 020813 (\# 12) are
excluded.  For comparison, the dotted line  shows the relation obtained by
Ghirlanda et al. (2004) based on their data base.}
\label{EcorEpeak}
\end{figure}

\newpage\clearpage

\begin{figure}[htp]
\includegraphics[width=1\textwidth]{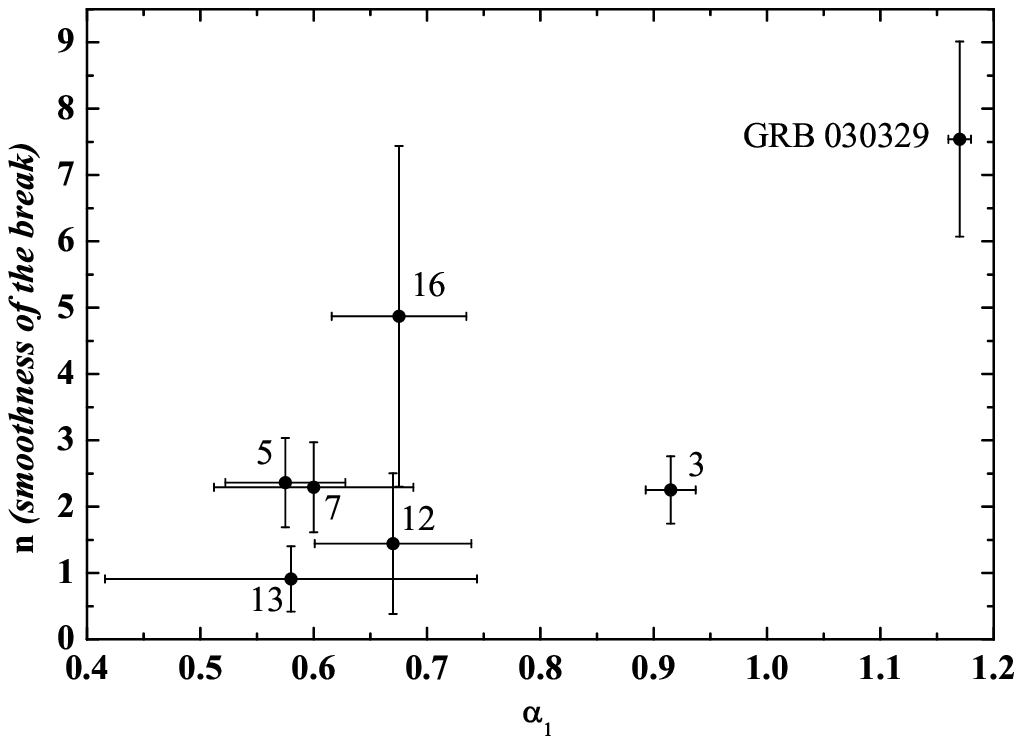}
\caption[]{The relation between the prebreak decay slope $\alpha_1$
and the smoothness of the break $n$ (Equation~\ref{mag1}). For all afterglows
not in this figure we had to fix the smoothness parameter to a relatively high
value of $n=10$ to get an acceptable fit (Table~\ref{tabres}).  All these fits
also find $\alpha_1>1$. In addition, we include here a special fit of the
afterglow of GRB 030329 as explained in the text (\S~\ref{correlations} and
Appendix \ref{individual}). This is the only fit where $\alpha_1>1$ and a
value for $n$ could be deduced.}
\label{a1n}
\end{figure}

\newpage\clearpage

\begin{figure}[htp]
\includegraphics[width=1\textwidth]{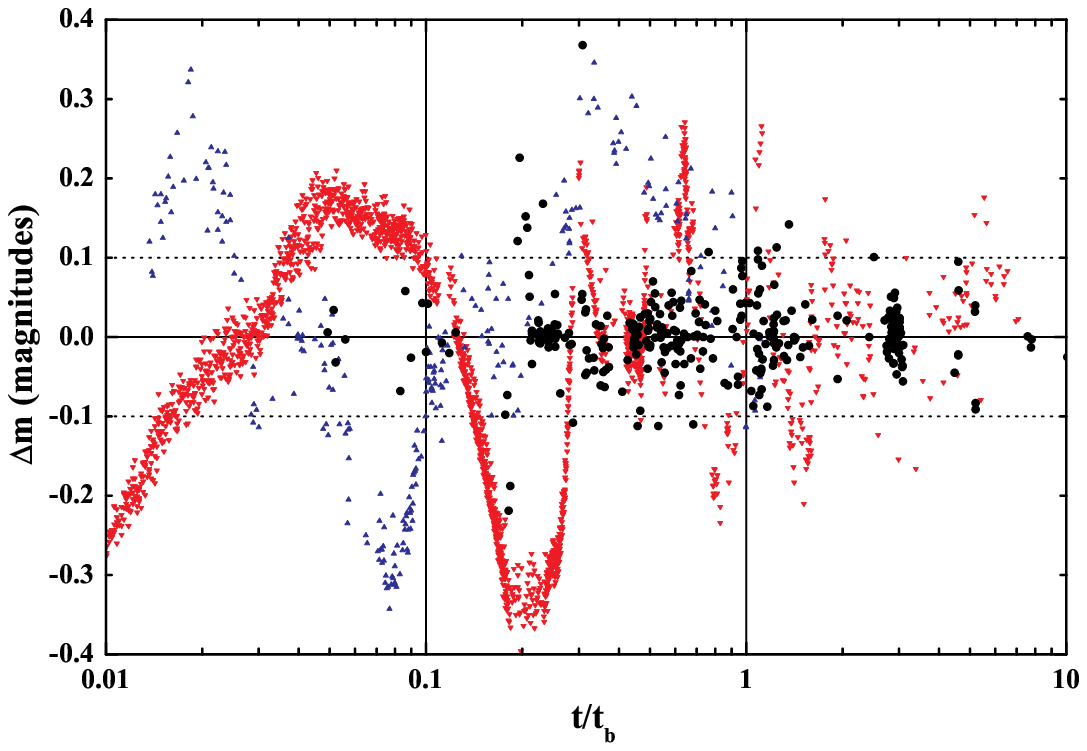}
\caption[]{
The residuals $\Delta m = m_{\rm observed} - m_{\rm fit}$ for all 16
afterglows in our sample (Table~\ref{tabres}, the large black points).
Included are only those data points where the reported magnitude error is less
or equal than 0.05 mag. The time axis is normalized to the corresponding break
time of the burst. The data indicate that any fine structure in the light
curves is on average less than $\pm$0.1 magnitudes with no evidence for
evolution. Note that the ratio $t/t_b$ is independent of redshift.  For
comparison, the residuals of GRB 021004 (upward pointing blue triangles) and
GRB 030329 (downward pointing red triangles) are plotted. The systematic
deviations from the Beuermann law are clearly seen, reaching almost 0.4
magnitudes. Furthermore, the initial fine structure of GRB 021004 and GRB
030329 is very similar; shifting the GRB 021004 light curve by a factor of 2.7
superposes this initial fine structure.}
\label{residuals}
\end{figure}

\end{document}